\documentclass[fleqn,usenatbib]{mnras}

\usepackage{newtxtext,newtxmath}

\usepackage{xcolor}
{}
\usepackage{cancel}
\usepackage{soul,xcolor}
\setstcolor{red}
\usepackage{hyperref}
\usepackage{threeparttable} 
\usepackage{booktabs}
\usepackage[T1]{fontenc}

\DeclareRobustCommand{\VAN}[3]{#2}
\let\VANthebibliography\thebibliography
\def\thebibliography{\DeclareRobustCommand{\VAN}[3]{##3}\VANthebibliography}

\usepackage{graphicx}	
\usepackage{amsmath}	
\usepackage{siunitx}
\usepackage{threeparttable}
\usepackage{comment}

\title[Magnetic Fields of the L328]{Magnetic fields on different spatial scales of the L328 cloud}

\author[Gupta et al.]{
Shivani Gupta,$^{1}$\thanks{E-mail: shivani.gupta@iiap.res.in}
Archana Soam,$^{1}$
Janik Karoly,$^{2,3}$
Chang Won Lee,$^{4,}$
Maheswar G.$^{1}$
\\
$^{1}$Indian Institute of Astrophysics, II Block, Koramangala, Bengaluru 560034, India \\
$^{2}$Jeremiah Horrocks Institute, University of Central Lancashire, Preston PR1 2HE, UK\\
$^{3}$Department of Physics and Astronomy, University College London, London WC1E 6BT, UK\\
$^{4}$Korea Astronomy and Space Science Institute (KASI)
776 Daedeokdae-ro, Yuseong-gu, Daejeon 34055, Republic of Korea\\
$^{5}$University of Science and Technology, Korea (UST), 217 Gajeong-ro, Yuseong-gu, Daejeon 34113, Republic of Korea}

\date{Accepted XXX. Received YYY; in original form ZZZ}

\begin{document}
\label{firstpage}
\pagerange{\pageref{firstpage}--\pageref{lastpage}}
\maketitle

\begin{abstract}

L328 core has three sub-cores S1, S2, and S3, among which the sub-core S2 contains L328-IRS, a Very Low Luminosity Object (VeLLO), which shows a CO bipolar outflow. Earlier investigations of L328 mapped cloud/envelope (parsec-scale) magnetic fields (B-fields). In this work, we used JCMT/POL-2 submillimeter (sub-mm) polarisation measurements at 850 $\mu$m to map core-scale B-fields in L328. The B-fields were found to be ordered and well-connected from cloud to core-scales, \textit{i.e.}, from parsec- to sub-parsec-scale. The connection in B-field geometry is shown using $Planck$ dust polarisation maps to trace large-scale B-fields, optical and near-infrared (NIR) polarisation observations to trace B-fields in the cloud and envelope, and 850 $\mu$m polarisation mapping core-scale field geometry. The core-scale B-field strength, estimated using the modified Davis-Chandrasekhar-Fermi relation{, was} found to be 50.5 $\pm$ 9.8 $\mu$G, which is $\sim$2.5 times higher than the envelope B-field strength found in previous studies. This indicates that B-fields are getting stronger on smaller (sub-parsec) scales. The mass-to-ﬂux ratio of 1.1 $\pm$ 0.2 suggests that the core is magnetically transcritical. The energy budget in the L328 core was also estimated, revealing that the gravitational, magnetic, and non-thermal kinetic energies were comparable with each other, while thermal energy was significantly lower.

\end{abstract}




\section{Introduction}

Magnetic field (B-field), turbulence, and gravity may play a crucial role in forming molecular clouds and stars \citep{2005Natur.438..332K}. B-fields are found to be very important (observationally and theoretically) in star formation but yet remain poorly measured. When the B-field dominates in a star-forming region, the core gradually condenses out of a magnetically subcritical background cloud \citep{1987ARA&A..25...23S, 1993prpl.conf..327M, 1999osps.conf..305M, 2003ApJ...599..363A}. 
This occurs when magnetic pressure dominates the gravitational pressure, causing the B-field lines to slowly redistribute mass. Over time, this gradual accumulation of mass leads to the formation of dense cores, eventually leading to a localized region where gravity can overcome magnetic support, allowing star formation to proceed.

At different spatial scales, the behavior of B-fields varies significantly in terms of their strength and morphology. On larger scales, greater than the size of molecular clouds (tens to hundreds of parsecs), B-fields can be more uniform, exerting a dominant influence on the dynamics of the interstellar medium and the formation of molecular clouds \citep{2012ARA&A..50...29C}. Within molecular clouds (a few parsecs), the B-field can start to show more structure and variability. On the core-scale (less than a parsec), the B-field becomes even more complex and tangled, with local variations in strength and orientation \citep{2012ARA&A..50...29C}. The strength of the B-field increases as we go from cloud-scale to core-scale.

The plane-of-the-sky B-field is measured by linear polarisation of dust {\citep{1949Sci...109..166H,1949ApJ...109..471H}} while the line of sight B-field can be obtained with the help of the Zeeman effect {\citep{crutcher1983magnetic, crutcher1999detection}}. To exhibit polarisation, dust grains should be asymmetrical in shape (which means elongated in any one direction); otherwise,  absorption remains uniform in all directions.  The minor axis of the dust grains is aligned with the B-field. So far, the Radiative Torque Alignment (RAT) is the most   {accepted} mechanism explaining the dust grain alignment in the ISM   {\citep{1976Ap&SS..43..291D,1996ApJ...470..551D,2007MNRAS.378..910L,2008MNRAS.388..117H, ALV2015,2022FrASS...9.3927T}}. Optical and Near-infrared (NIR) polarisation is a result of dichroism or selective absorption of the background starlight by dust grains \citep{2003JQSRT..79..881L}. In contrast, longer wavelength (far-infrared (FIR) to millimeter (mm)) polarisation is a result of thermal emission from dust grains. At longer {wavelengths}, we obtain the B-field orientation by rotating the polarisation angle by 90$^\circ$ because the thermal emission from the dust grains is polarized along the major axes of the aligned grains. Polarisation observations at different wavelengths are used to trace B-fields at different extinction ($   {A_{\rm V}}$) levels. Optical traces $   {A_{\rm V}} \sim$ 1–10 mag \citep{1995ApJ...448..748G}, NIR traces $   {A_{\rm V}}$ $\sim$ 10–20 mag, sub-mm traces $   {A_{\rm V}}$ $\sim$ 20-50 mag \citep{2015A&A...574C...4A} and mm traces even ${A_{\rm V}}$ > 50 mag {\citep{1995ApJ...448..346T,1999sf99.proc..212T}}. At sub-mm/mm wavelength, since the dust grains are relatively cooler, we trace the denser parts (high $   {A_{\rm V}}$ regions) of the cloud, \textit{i.e.}, cores where the dust grains are shielded \citep{2009MNRAS.398..394W}.

In protostars, the outflows can impact their surrounding environment by inducing turbulence \citep{2015A&A...573A..34S}, disturbing any initial alignment between the core and the envelope B-fields. Very Low Luminosity Objects (VeLLOs) are similar to typical protostars with the low bolometric temperature (< 650 K) and very low luminosity, less than $   {L}$ $<$ 0.1 $   {L_{\odot}}$ and weaker energetic outflows compared to the typical Class 0/I outflows associated with low-mass stars \citep{2002A&A...393..927B,2004A&A...426..503W,2006ApJ...649L..37B,2011ApJ...743..201P}. With the low luminosity and less energetic outflow, we expect VeLLOs to induce less turbulent effects on their surroundings. This would enable the regions to retain the initial condition that may have existed before the beginning of star formation, revealing primordial B-ﬁelds \citep{2015A&A...573A..34S,soam2015first}. 


{Lynds Dark Nebulae (LDN)}\,328 (hereafter L328), initially classified as a starless dense core located at a distance of $\sim$217 pc \citep{maheswar2011distances}, was found to harbor {a VeLLO} in one of the three sub-cores by the $Spitzer$ telescope \citep{2009ApJ...693.1290L}. Of the three {sub-cores, S1, S2, and S3, the S2 sub-core} harbours the VeLLO L328-IRS {(Infrared Source)} \citep{lee2013early}. The presence of an outflow was detected by CO (2-1) line emission in \cite{lee2013early}, and its more detailed structures were further studied with ALMA {(Atacama Large Millimeter/submillimeter Array)} observations \citep{lee2018high}. The 1.3 mm continuum ALMA observations confirmed the detection of a disk around L328-IRS with the mass accretion rate of 8.9$\times$10$^{-7}$ $   {M_{\odot}}$ yr$^{-1}$. The disk is fitted with a Keplerian model from 60 to 130 AU, confirming its rotation \citep{lee2018high}. The position of L328-IRS {(S2 sub-core)}, L328-S1 {(S1 sub-core)} and L328-S3 {(S3 sub-core)} are at ($\alpha,\delta$)$_{J2000}$ = (18$^{h}16^m59^s$.50, -18°02'30.5"), ($\alpha,\delta$)$_{J2000}$ = (18$^{h}16^m59^s$.55, -18°02'06.5"), and ($\alpha,\delta$)$_{J2000}$ = (18$^{h}17^m00^s$.88, -18°02'09.0"), respectively.

L328, along with L323 and L331, forms a system of three cometary globules that are found near the OB association stars. All three clouds (dark nebula) show a similar orientation of head-tail morphology, suggesting the same ionising source. This system has three ionising stars, which are close to B-type stars named HD16832, HD168675, and HD167863, located within 2°. The ionisation source produces shock (by ionisation heating) that converges into the cloud, resulting in compression and formation of single and multiple cores \citep{2023MNRAS.524.1219K}.

Earlier studies have analyzed the B-field morphology in L328 using ${Planck}$, optical, and NIR polarisation measurements {\citep{2015A&A...573A..34S, soam2015first,  2016A&A...594A...1P, 2016A&A...594A..26P,2023MNRAS.524.1219K}}. {This work extends the investigation by including core-scale B-field morphology at 850 ${\mu}$m, illustrating how the B-fields are connected across different wavelengths and spatial scales.}


The structure of the paper is as follows: in {Section \ref{sec2}}, we present the observations and data reduction. {In Section \ref{result}}, we include results, {analysis}, and the discussion. {Section \ref{summary}} summarizes our work.

\section{Observations and data reduction}\label{sec2}

The observations were conducted with SCUBA-2/POL-2 at 850 $\mu$m in 2018 March (M18AP033; PI: Archana Soam) and in 2019 May and June (M19AP014; PI: Chang Won Lee) using the Daisy-map mode of the JCMT \citep{holland2013}, optimized for POL-2 observations \citep{friberg2016}. The POL-2 polarimeter, which consists of a ﬁxed polarizer and a half-wave plate rotating at a frequency of 2 Hz, is placed in the optical path of the SCUBA-2 camera. The weather conditions during observations were split between $\tau_{225}$<0.05 and 0.05<$\tau_{225}$ <0.08, where $\tau_{225}$ is the atmospheric opacity at 225 GHz {as measured by a radiometer 225 GHz at the Sub-Millimeter Array (SMA). The atmospheric opacity at 225 GHz can then be converted to the atmospheric opacity at 353 GHz \citep[850~$\mu$m;][]{holland2013,mairs2021}, but the weather bands for JCMT are left in terms of values relating to $\tau_{225}$.}
The total integration time for a single ﬁeld was $\sim$ 31 minutes, and there were 17 repeats for a total on-source integration time of $\approx$8.8~hrs. SCUBA-2/POL-2 simultaneously collects data at 450 and 850 $\mu$m with effective FWHM beam sizes of 9\farcs6 and 14\farcs1, respectively. For this work, we focus exclusively on the 850 $\mu$m data. This observing mode is based on the SCUBA-2 constant velocity Daisy scan pattern, but modiﬁed to have a slower scan speed ({\textit{i.e.}}, 8"s$^{-1}$ compared to the original 155" s$^{-1}$) to obtain sufﬁcient on-sky data to measure the Stokes Q and U values accurately at every point of the map \citep{holland2013}. The integration time decreases toward the edges of the map, which consequently leads to an increase in the rms noise levels. This scan pattern gives a 3$\arcmin$ diameter central region with uniform noise characterization, though this has been shown to extend out to a diameter of 6$\arcmin$ \citep{doris2021}.

To reduce the data, we used the STARLINK/SMURF package \textit{pol2map} speciﬁcally developed for {sub-mm} data obtained with the JCMT \citep{chapin2013,currie2014}. The details of the data reduction procedure are presented in \cite{Wang_2019} and we will only summarize the relevant steps here. First, the raw bolometer time streams are converted into Stokes {\textit{I}, \textit{Q}, and \textit{U}} time streams at a sampling rate of a full half-wave plate rotation through the process \textit{calcqu}. An initial Stokes \textit{I} map is created from the Stokes \textit{I} time streams using the command \textit{makemap} \citep{chapin2013}. Then, the ﬁnal Stokes \textit{I}, \textit{Q}, and \textit{U} maps were obtained by running \textit{pol2map} a second time. This second {iteration} uses the initial Stokes \textit{I} map to mask the areas with astronomical signal and then runs a version of \textit{makemap} called `skyloop' which runs the map maker on the observations in parallel rather than one-by-one. We set the `mapvar' parameter to be used and so the final errors in the map were estimated from the spread of errors across the 17 observations. A comprehensive review of the map-making process and the removal of sky background and other sources of noise is given in \citet{chapin2013}. In summary, the sky background is removed iteratively and is treated as a common-mode signal across the bolometers. The map-maker also models other components which are iteratively removed until only the astronomical signal remains. We also correct for instrumental {polarisation} (IP) in the Stokes Q and U maps based on the final Stokes I map and the “August 2019” IP {polarisation} model\footnote{\url{https://www.eaobservatory.org/jcmt/2019/08/new-ip-models-for-pol2-data/}}. Once the final Stokes \textit{I}, \textit{Q}, and \textit{U} maps were made, we made a polarisation vector catalog by binning up from the 4" pixel size to 12", which approximates the beam size 14\farcs1. This helps in reducing the number of vectors by combining vectors within each 12" pixel and also decreasing the noise level. Specifically, for plotting, we selected vectors with an intensity-to-error ratio ($   {I/\delta I}$) > 10 and a polarisation-to-error ratio ($   {P/\delta P}$) > 2.

In order to convert the native map units of pW to astronomical units, a flux calibration factor (FCF) of 497.5397 Jy/beam/pW was used \citep{mairs2021}, multiplied by a factor of 1.35 to account for POL-2 being inserted into the beam. The peak values of total and polarized intensities are found to be $\sim$100 mJy beam$^{-1}$ and $\sim$11 mJy beam$^{-1}$, respectively. The rms noise of the background region in the Stokes \textit{I}, \textit{Q}, and \textit{U} maps is measured to be $\sim$6.26 mJy/beam, 5.27 mJy/beam, and 5.75 mJy/beam, respectively. These values were determined by selecting a region about 1' from the center of each corresponding map, where the signal remained relatively constant. This region is relatively ﬂat, exhibits moderate unpolarisation, and low emission, and is distanced from the brightest region in the corresponding maps.

The values for the debiased degree of polarisation $   {P}$ were calculated using the modified asymptotic estimator \citep{plaszcz2014,montier2015}
\begin{equation}
P = \frac{1}{I} [PI - 0.5\sigma^{2}(1-e^{-(PI/\sigma)^{2}}/PI ] \,\, ,
\end{equation}
\noindent
where \textit{I}, \textit{Q}, and \textit{U} are the Stokes parameters, and $\sigma^{2}=(Q^{2}\sigma_{Q}^{2}+U^{2}\sigma_{U}^{2})/(Q^{2}+U^{2})$ where $\delta Q$, and $\delta U$ are the uncertainties for Stokes \textit{Q} and \textit{U}

The polarisation position angles $\theta$, measured from north through east on the plane of the sky, were calculated using the relation 
\begin{equation}
{\theta = \frac{1}{2}{\rm tan}^{-1}\frac{U}{Q}} \, ,
\label{eq:theta}
\end{equation}
\noindent
and the corresponding uncertainties in $\theta$ were calculated using
\begin{equation}
   {\delta\theta = \frac{1}{2}\frac{\sqrt{Q^2\delta U^2+ U^2\delta Q^2}}{(Q^2+U^2)}} \,\, .
\label{eq:dtheta}
\end{equation}

The plane-of-sky orientation of the {B-field} is inferred by rotating the polarisation angles by 90$^{\circ}$. As mentioned in the Introduction, this {rotation is due to the fact} that the polarisation is caused by elongated dust grains aligned perpendicular to the {B-field} \citep[see][and references therein]{ALV2015}.
The polarisation results from JCMT/POL-2 are given in table \ref{tab:vector}.

\section{Results and discussion}\label{result}

\begin{figure*}
\begin{center}
\resizebox{16.0cm}{19.0cm}{\includegraphics{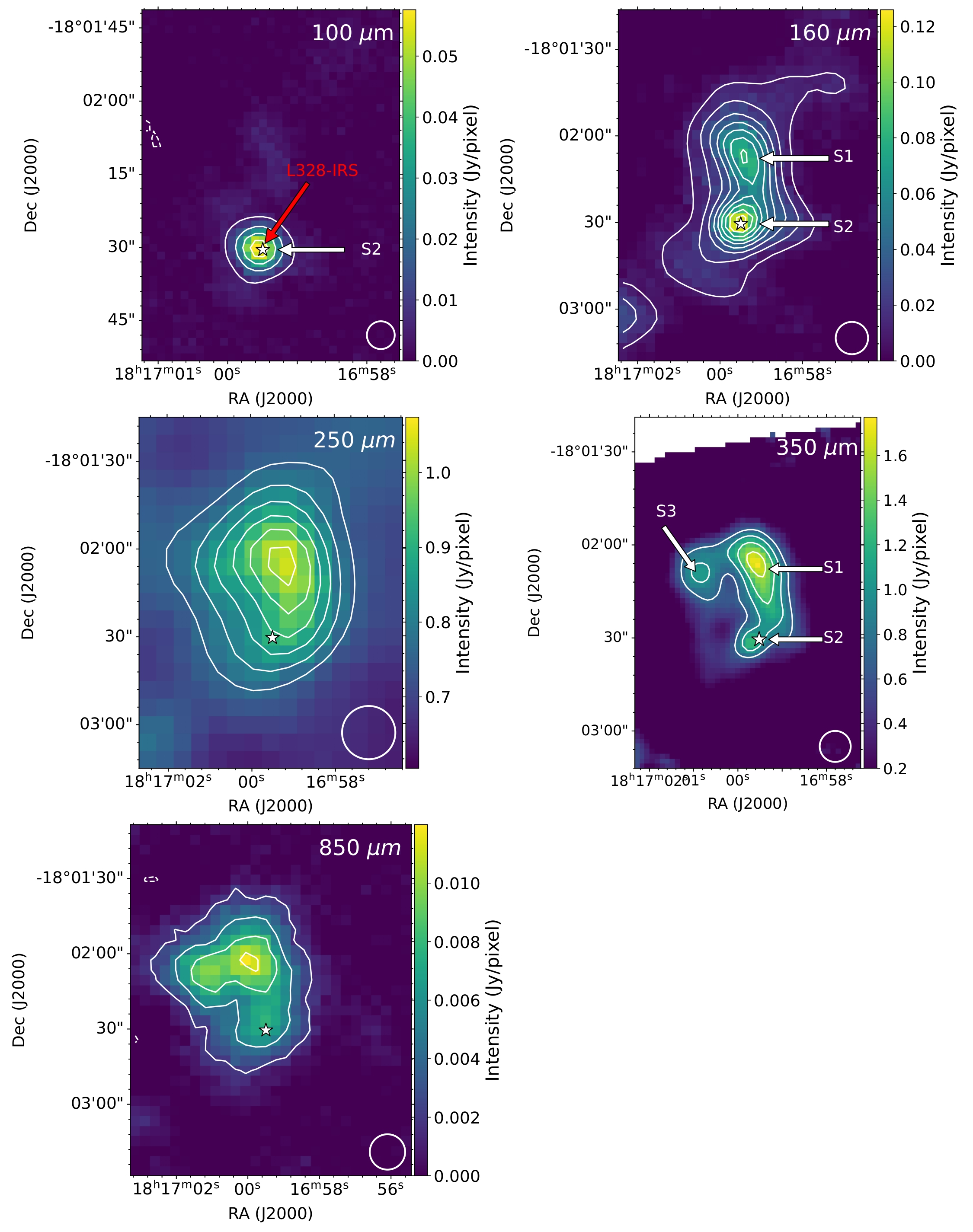}}
\caption{The dust emission maps with total intensity contours in 100, 160 and 250 $\mu$m wavelengths from \textit{Herschel}/PACS and SPIRE data archive. The 350 $\mu$m emission map is from {SHARC2}. The 850 $\mu$m is mapped with JCMT/SCUBA-2 in this work. The star symbol shows the position of L328-IRS. The white circle in the bottom right corners is the beam size in each panel. {The contour levels, in Jy/pixel, are drawn at [0.006, 0.017, 0.028] for 100~$\mu$m, [0.01, 0.02, 0.03, 0.04, 0.05, 0.06, 0.07] for 160~$\mu$m, 
[0.7, 0.8, 0.85, 0.9, 0.94] for 250~$\mu$m, [0.8, 1.2, 1.6, 2.0] for 350~$\mu$m, and [0.003, 0.005, 0.007, 0.009] for 850~$\mu$m.
}}\label{Fig:contour5}
\end{center}
\end{figure*}

\begin{table*}
\begin{center}
	\caption{Summary of data set for {photometry of the S2 sub-core}.}
	\label{tab:photometry}
	\begin{tabular}{lccccc} 
		\hline
		Wavelength  & Instrument & S2 Flux Density   & $\sigma$  & Aperture  & Beamsize\\ 
		   ($\mu$m)  &            &    (mJy)  &    (mJy)  &    (") &    (")  \\ 
		\hline
		70   & \textit{Spitzer} c2d/IRAC and MIPS & 281 & 41 &  29.7& 5.7 \\
            100  & \textit{Herschel}/PACS & {1366} & {16} & 14.2 & 7.1\\ 
		160  & \textit{Herschel}/PACS & {2470} & {28.8}& {20.0} &11.2\\ 
		250  & \textit{Herschel}/SPIRE & - & - & -& 18.2\\
		350  & {CSO/SHARC2} & 3200 & 950 & 20.0 &10\\
		850  & SCUBA-2/POL-2 & 128.5 & 6.25 & 20.0 & 14.1\\ 
		1200 & IRAM &70 & 80 & 20.0 &\\
		\hline 
	\end{tabular}
    \begin{tablenotes}
        {Note:} The values for 70, 350 and 1200 $\mu$m are taken from \cite{lee2009spitzer} and beam size at 350 $\mu$m from \cite{suresh2016catalog}. At 250 $\mu$m {the} table does not include photometric data as emission from S2 sub-core is not distinguishable from other two sub-cores. 
    \end{tablenotes}
	
\end{center}
\end{table*}

\begin{figure}
\resizebox{9.0cm}{7.0cm}{\includegraphics{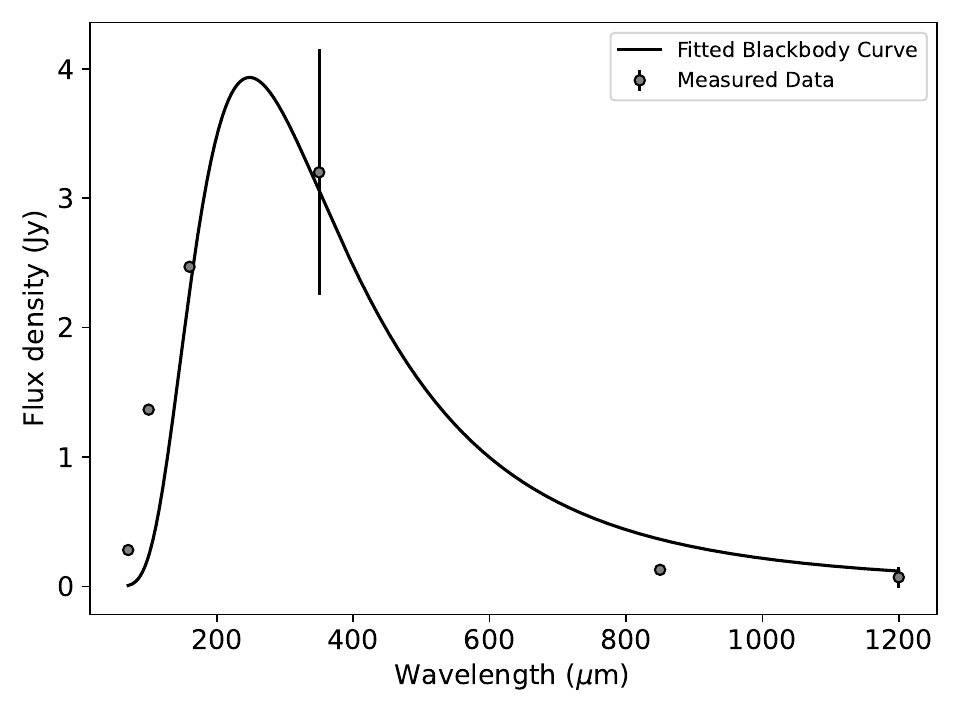}}
\caption{ {The photometric data of sub-core S2} (given in Table \ref{tab:photometry}) is fitted with a black-body curve.}\label{Fig:sed_fit}
\end{figure}

\begin{figure*}
\begin{center}
\resizebox{18.5cm}{16cm}{\includegraphics{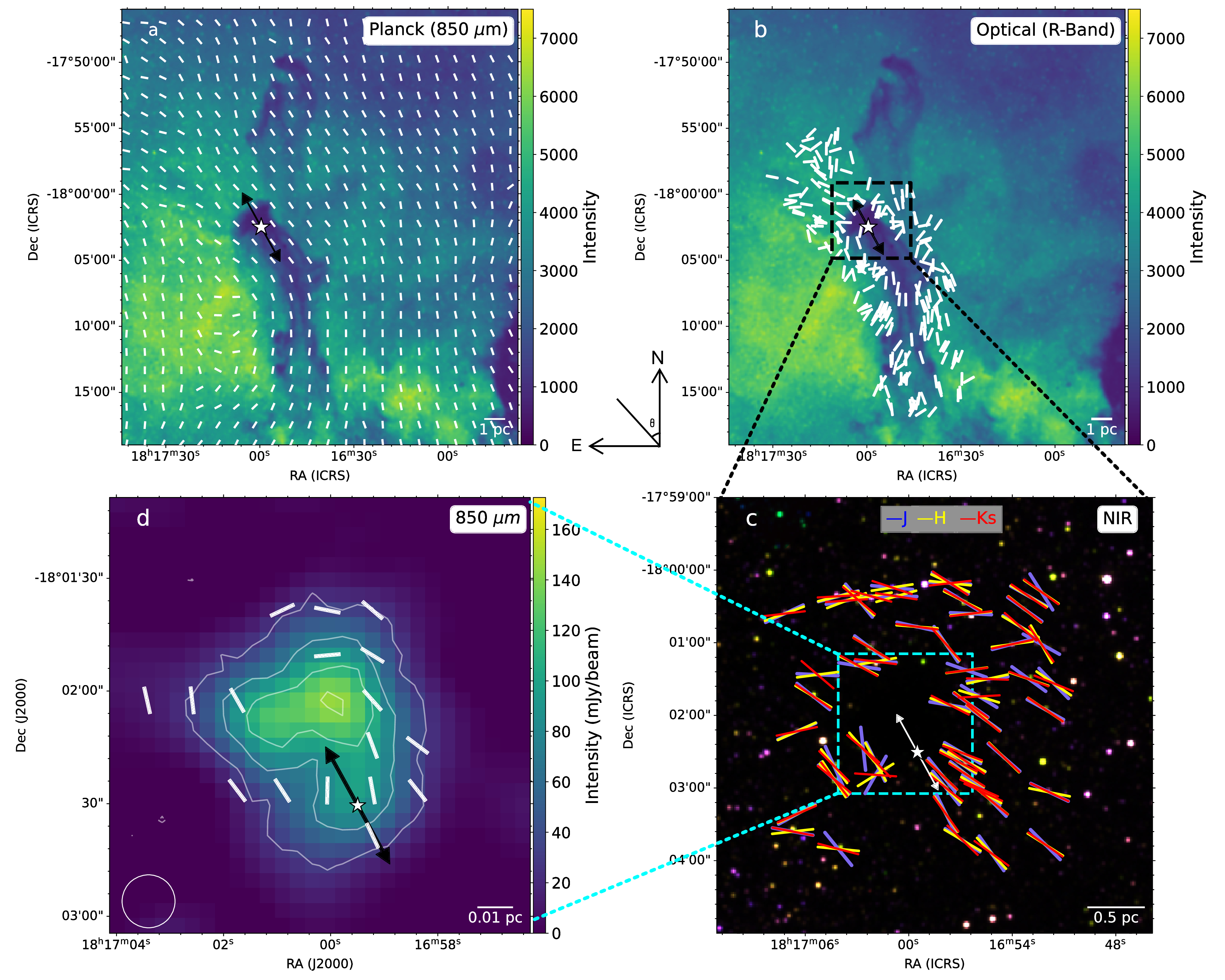}}
\caption{{\bf Panel (a):} Morphology of B-fields obtained from \textit{Planck} 850 $\mu$m dust polarisation observations overplotted \citep{2016A&A...594A...1P,2016A&A...594A..26P} on the continuum-subtracted H$\alpha$ image of the L328 region.  {\bf Panel (b):} The B-fields mapped with optical R-band (0.63 $\mu$m) observations by \citet{2015A&A...573A..34S} overplotted on the same image as panel(a).
{\bf Panel (c):}  The polarisation vectors are shown in blue (J), yellow (H), and red (Ks) are overplotted on the color-composite image. 
{\bf Panel (d):} The B-field morphology obtained from 850 $\mu$m dust polarisation observations and contours of intensity overplotted on 850 $\mu$m dust emission continuum map of L328 core. The beam size of 14\farcs1 is shown with open circle in bottom left corner. The location of L328-IRS in all panels is shown with star symbol and its outflow direction with a double-headed arrow as it has bipolar outflow. The lengths of line-segments are normalized and independent of fraction of polarisation.} 
\label{Fig:magnetic field}
\end{center}
\end{figure*}

\begin{figure*}
\centering
\resizebox{18.5cm}{16cm}{\includegraphics{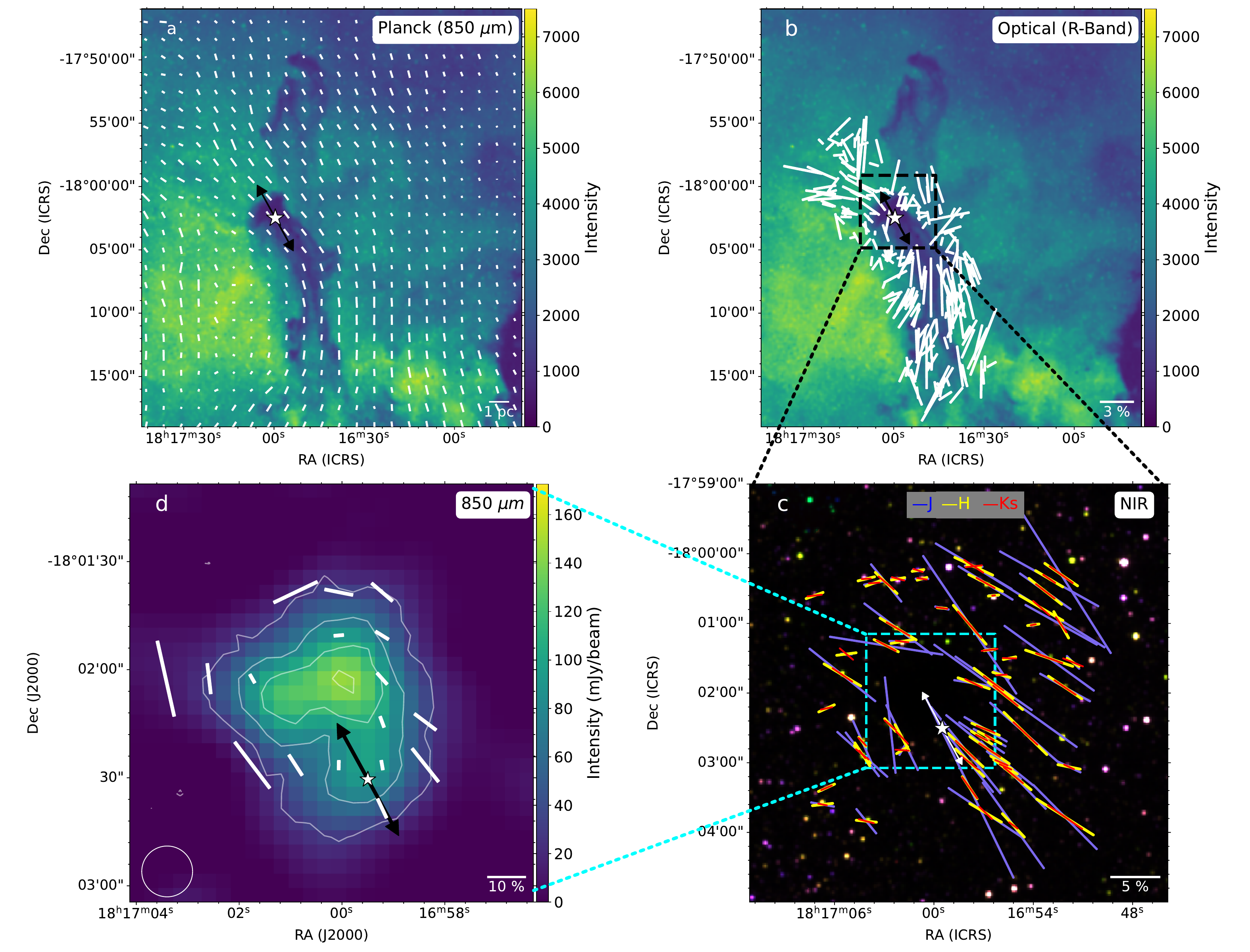}}
\caption{In addition to figure \ref{Fig:magnetic field}, each panel in this figure depicts polarisation vectors of varying lengths, representing different polarisation percentages. The length of each bar in the bottom right corner indicates the corresponding polarisation percentage.} \label{Fig:manetic+pol}
\end{figure*}


\begin{figure*}
\begin{center}
\resizebox{14.0cm}{17.0cm}{\includegraphics{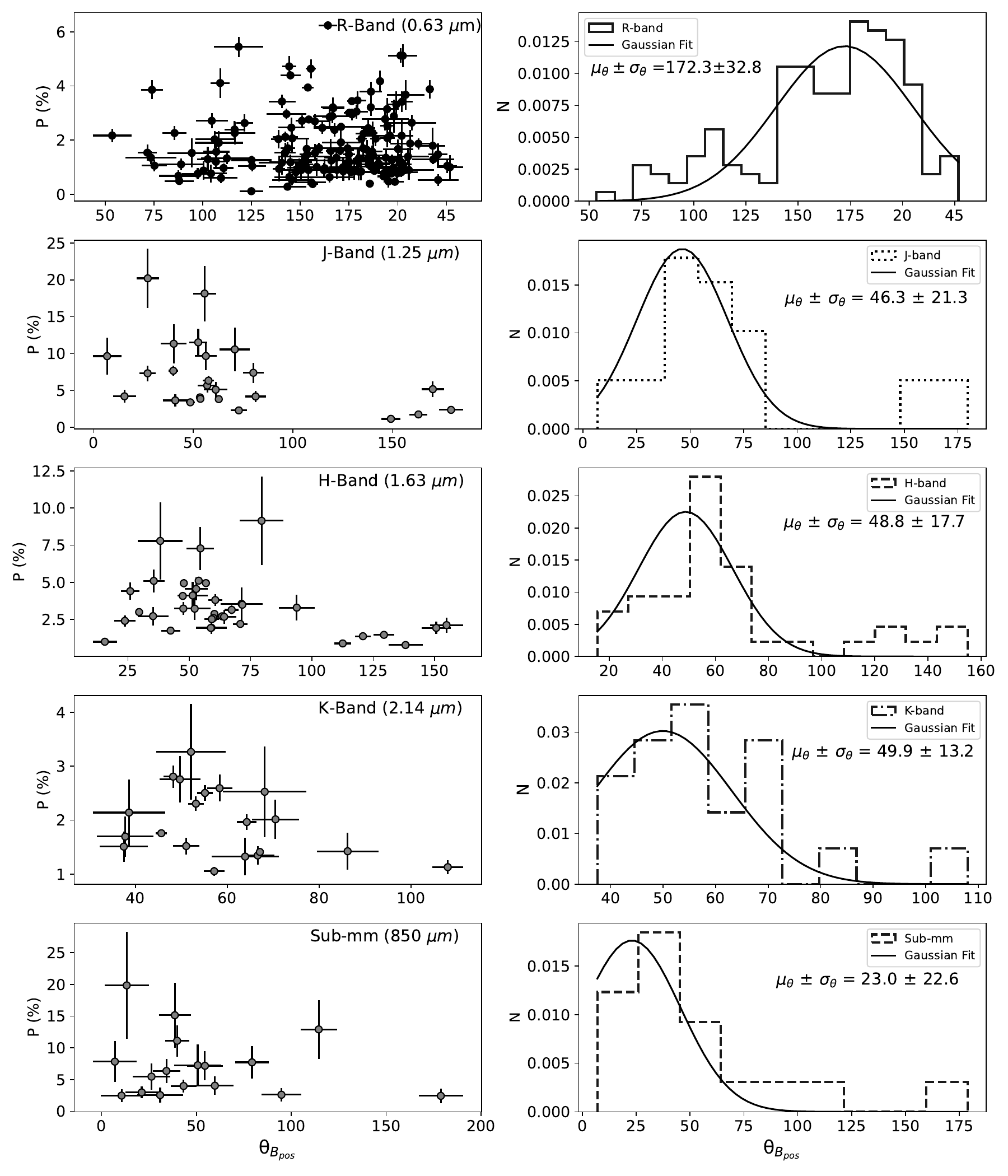}}
\caption{The left panel shows the distribution of the degree of polarisation against the position angle of the {B-field} in {R}, J, H, and K bands, and at sub-mm wavelength (850 $\mu$m) in the L328 region. The right panel shows the mean and variance of Gaussian-fitted histograms of position angle of {B-field} in the respective bands.\label{Fig:JHK}}
\end{center}
\end{figure*}

\begin{figure*}
\begin{center}
\resizebox{13cm}{18.0cm}{\includegraphics{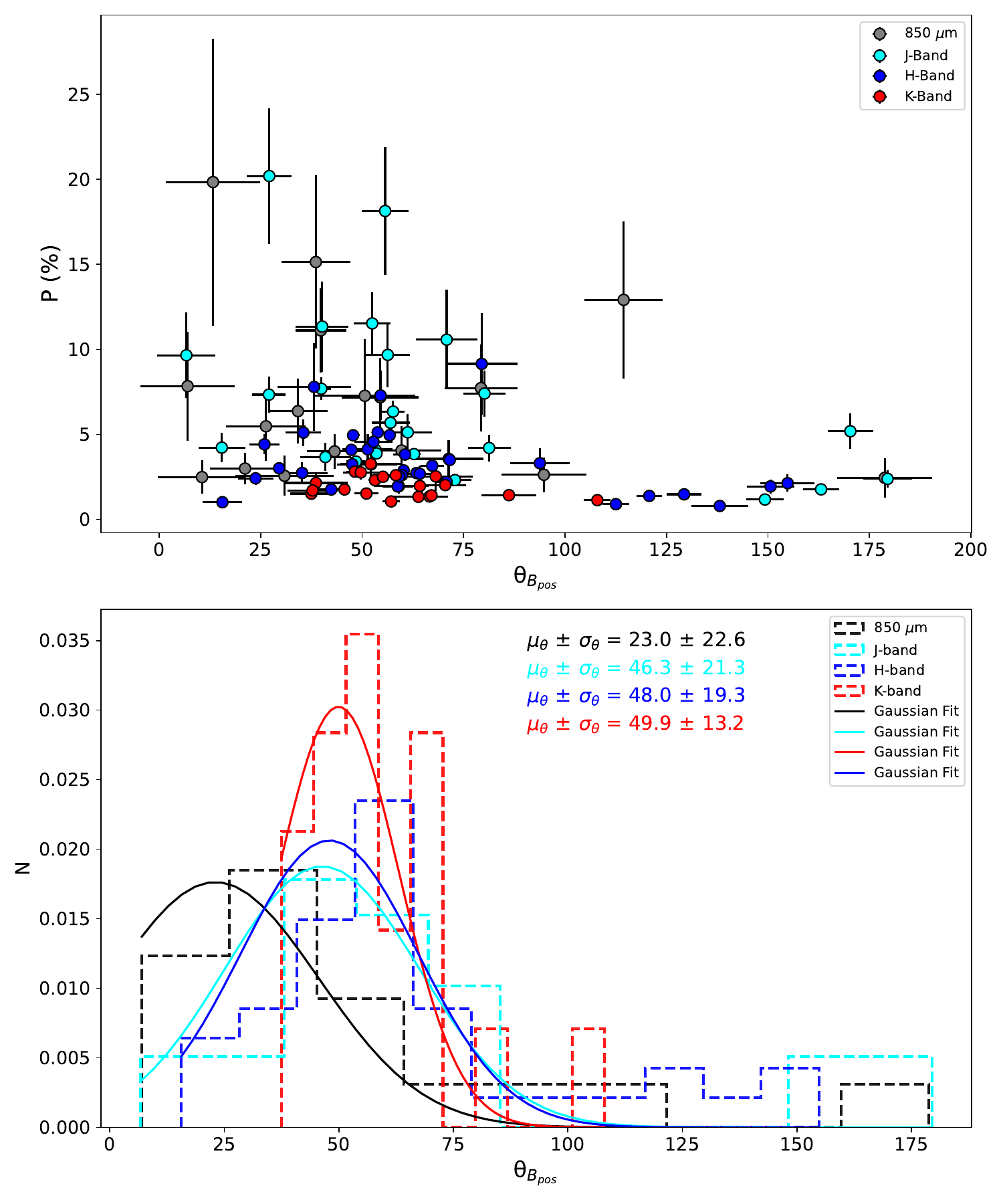}}
\caption{ The upper and lower panels show a comparative study at J, H, K bands, and sub-mm wavelengths by plotting the distribution of the degree of polarisation vs. the position angle of the {B-field} in the upper panel and the mean and variance of Gaussian-fitted histograms of the position angle of the {B-field} in the lower panel. }
\label{fig:com_JHK}
\end{center}
\end{figure*}

\subsection{Structure and Kinematics of L328}

In \cite{2007AJ....133.1560W}, it is shown that at {350 $\mu$m}, {the} L328 core has three sub-cores named S1, S2, and S3 forming at the same time as shown in Figure \ref{Fig:contour5}.
Further, in \cite{lee2009spitzer}, by using $Spitzer$ data, they discovered L328-IRS in {the} L328 cloud, which earlier was thought to be starless. \cite{lee2013early} confirmed that the L328-IRS is associated with sub-core S2 by molecular observations using $^{13}$CO and N$_2$H$^{+}$ which are highly {broadened} towards sub-core S2 whereas S1 and S3 are starless cores. At sub-parsec scales, L328-IRS shows the bipolar outflow in North-East and South-West direction detected in CO \citep{lee2013early}. Infall asymmetry in N$_2$H$^{+}$(1-0), CO(3-2) and HCN(1-0) lines shows inward motion towards L328-IRS. The L328 core is located at the head of L328 cloud as shown in panel (b) of Figure \ref{Fig:magnetic field} in the R-band DSS image \citep{2015A&A...573A..34S}. L328-IRS is further studied by \cite{lee2018high} using ALMA observations at 1.3 mm continuum (Band 6) and $^{12}$CO, $^{13}$CO, and C$^{18}$O {(2–1)} molecular lines. They found a rotating disk around L328-IRS whose deconvolved size is $\sim$ 87 $\times$ 37~AU and {a more detailed X-shaped bipolar outflow} with opening angle of 92$^{\circ}$ and inclination angle of 66$^{\circ}$ (angle between line of sight and the outflow axis).

{For the structure of the L328 core}, we explored $Hershel$ data at 100, 160, 250, and 350 $\mu$m wavelengths and JCMT data at 850 $\mu$m. In Figure \ref{Fig:contour5}, we plotted intensity contours at 5 different wavelengths in the L328 core. At 100 $\mu$m, only sub-core S2 is visible. Moving to 160 $\mu$m, {sub-core S1 becomes visible}, and the L328 core exhibits an elongated structure in North-West (a narrow emission feature) and South-East direction (a broad emission feature). At 250 $\mu$m, the emission is from sub-core S1, S2, and S3, though it may not be resolved due to its beam size. At 350 $\mu$m, all three sub-cores are visible and labelled.{ The emission from sub-cores S1 and S3 initiates at longer wavelengths compared to sub-core S2 (see Figure \ref{Fig:contour5}), indicating that sub-core S2, which contains L328-IRS, is hotter than the other two starless sub-cores, S1 and S3.} This flux variation at different wavelengths is further explored through spectral energy distribution (SED) fitting in the next section.

\subsection{{Spectral energy distribution fitting}}
We have the L328 emission maps in various wavelengths which give us an opportunity to compare fluxes and brightness of the core in different apertures. For the photometry, we used the package {\textit{photutils}} in python \citep{larry_bradley_2023_7946442}. For sub-core S2, we used the coordinate of L328-IRS as the centre of circular aperture and we decided the size of aperture in such a way that there would be no contribution from S1 and S3 sub-cores. {The SED fitting was done using following formula for blackbody emission}

\begin{equation}
   {B_{\nu}(T_{\rm d})=\frac{2h v^{3}}{c^{2}} \frac{1}{e^{hv/k_{B}T_{\rm d}}-1}} \,\,,
\end{equation}
\begin{equation} 
{{S}_{\nu}=\Omega\,B_{\nu}(   {T_{\rm d}})} \,\,, \label{eq:flux_density}
\end{equation}
where $   {B_{\nu}(T_{\rm d})}$ is the Planck function, $   {S_\nu}$ is flux density at the frequency $\nu$ and $\Omega$ is aperture size in solid angle. The photometric data {used for fitting} is given in Table \ref{tab:photometry}. The temperature for the S2 sub-core is found to be 11.5 K using SED fitting equation \eqref{eq:flux_density}, as shown in Figure \ref{Fig:sed_fit}. { This temperature provides the best fit with the least ${\chi^2}$ between observed and estimated flux densities.}
This temperature is in agreement with the value of $   {T_{\rm d}}$ = 16 K for the sub-core S2 in \cite{lee2013early}. Similarly, we fit an SED for sub-core S1, the temperature found to be 10 K. So, for further calculations of L328-core, we have taken the temperature as $   {T_{\rm d}}$ = 11.5 K.


\subsection{Core mass estimation}{\label{mass_est}}
We estimated the mass of the core using 850 $\mu$m dust continuum observations used in this work. The mass of the core is estimated using relation



\begin{equation}
M_{\rm core} =    {D} \frac{   {S_{\nu}}    {d}^{2}}{   {B_{\nu}(T_{\rm d})}    {k_{\nu}}} \,\,,
\end{equation}
  {where} $   {S_{\nu}}$~(Jy) is the ﬂux density at 850 $\mu$m, $   {d}$ is the distance of the core, $   {k_{\nu}}$ is the dust opacity, which is adopted as 1.85 cm$^2$ g$^{-1}$ from \cite{ossenkopf1994dust}, and $   {D}$ is the gas-to-dust mass ratio,   {taken as 100}.

We used a temperature of 11.5 K and 10 K for sub-core S2 and S1, respectively, as determined from the SED fitting section.  We performed photometry at 850 $\mu$m for the whole L328 core and S1 sub-core with aperture sizes of 72" and 30", respectively, resulting in total flux densities of 1.12 Jy and 338 mJy. Based on these calculations, the masses of L328 core, S1 sub-core, and S2 sub-core are found to be {0.69} $   {M_{\odot}}$, {0.34} $   {M_{\odot}}$, and {0.08} $   {M_{\odot}}$, respectively. \cite{lee2013early} reported the mass of {L328 core, S1 sub-core, and }sub-core S2 as {1.3} $   {M_{\odot}}$, {0.7} $   {M_{\odot}}$, and 0.09 $   {M_{\odot}}$ using 350 $\mu$m dust continuum. Usually, mass estimations using dust emission suffer from high uncertainty, at least by a factor of 2, due to uncertain dust opacity. This uncertainty arises because different models assume varying values for $   {k_{\nu}}$, which depend on factors such as grain size, composition, and temperature.

\subsection{Magnetic field morphology}
We investigated the B-field morphology in L328 cloud using existing {${Planck}$,} {optical, and NIR} polarisation measurements  {\citep{2015A&A...573A..34S, soam2015first,  2016A&A...594A...1P, 2016A&A...594A..26P, 2023MNRAS.524.1219K}} 
and the core region using sub-mm polarisation observations of this work. The B-field morphology at {different spatial scales, obsereved using different wavelengths,} is shown in Figure \ref{Fig:magnetic field}. The polarisation vectors (line-segments) in this figure are normalised to the same length, {\textit{i.e.}}, length is independent of the polarisation values associated with the vectors. In panel (a), $Planck$ polarisation and in panel (b) the optical polarisation vectors are overplotted on the same continuum-subtracted H$\alpha$ image adapted from \cite{soam2015first}. At optical wavelengths, L328 is opaque, and thus no background stars are seen toward L328. Panel (c) shows the NIR polarisation vectors and zooms in on a black-highlighted square box ($\sim 6^{\prime}$ $\times$ $5.8^{\prime}$) of panel (b). Panel (c) shows the background image taken from 2MASS survey where J, H, and {Ks} Band traces 1.25, 1.63, and 2.14 $\mu$m, respectively. Panel (d) further zooms in on a cyan-highlighted square box ($\sim 1.9^{\prime}$ $\times$ $1.9^{\prime}$) from Panel (c) on a sub-parsec scale, specifically tracing B-field in the core. The estimated area of L328 core from the last overlaid contour is $\approx$ 13124 $\times$ 14530~AU, or if considered a sphere, then its radius is $\sim$7800 AU. 

The large scale B-fields seen with $Planck$ polarisation are found parallel to the whole cloud L328 major axis, overall tracing the B-field orientation in  {Northeast-Southwest} direction. The optical polarisation observations trace the periphery of the cloud. Optical polarisation vectors seem to trace random {orientation} but maintaining overall orientation to be in  {Northeast-Southwest} direction. The NIR region traces denser regions around the core within the cloud. In panel (c) of Figure \ref{Fig:magnetic field}, there are three sets of polarisation vectors denoted with different colors— blue for J-band, yellow for H-band, and red for Ks-band polarisation vectors. The data is adopted from \cite{soam2015first}. The orientation of B-field shown by these J, H, and Ks bands is approximately similarly structured, but the degree of polarisation is different for these, as we can see in panel (c) of Figure \ref{Fig:manetic+pol}.  
At the same spatial location on the sky, the J-band polarisation vectors (blue) are significantly longer than those of the H-band (yellow) and Ks-band (red), indicating a higher degree of polarisation (\%) in the J-band than the others. Similarly, the H-band vectors are longer than the Ks-band vectors, suggesting a higher polarisation (\%) in the H-band than in Ks-band.
In panel (d) of Figure \ref{Fig:magnetic field}, the 850 $\mu$m polarisation measurements are used to trace core-scale B-fields. These B-fields morphology resemble those observed at envelope scale, showing the connection between the core and envelope scale. Furthermore, a clear bend in the field lines can be seen on the upper-right shoulder of the core. It appears that field lines warp the core outer boundary. The vectors in the lower part of the core are found parallel to the outflow axis.

 The mean and variance of $\theta_{{B_{\rm pos}}}$ are analysed with the help of Gaussian fitted histogram as shown in right side of Figure {\ref{Fig:JHK}}. The mean $\theta_{{B_{\rm pos}}}$ for optical is 172.3° with variance 32.8° representing a number of polarisation vectors parallel to the cloud. However, \cite{2023MNRAS.524.1219K}, used the histogram of relative orientation (HRO) technique to more precisely determine the orientation of B-field with respect to cloud morphology of L328 at optical wavelength,
 which suggests that the B-field is preferentially perpendicular to the cloud structure, something that can be seen in Figure \ref{Fig:magnetic field}(b).
 The mean $\theta_{{B_{\rm pos}}}$ for J, H, and Ks {is $\sim$45° for the three bands}, which represents  {Northeast-Southwest} direction. The mean $\theta_{{B_{\rm pos}}}$ for sub-mm emission is 23° with variance 22.6°. The angular offset between the mean B-field orientation and the orientation of the outflow axis (\textit{i.e.}, 30°) is measured as 16°, 19°, and 20° in the near-IR for J, H, and Ks bands, respectively. In the sub-mm range, this offset is found to be 7°, {though it has low statistical confidence due to limited number of vectors at 850 ${\mu}$m.}

 {There is a hint of overall B-fields along { {Northeast-Southwest} direction at ${Planck}$, NIR, and 850 ${\mu}$m,} suggesting a clear connection between {cloud to} core scale field lines.}

 We examined the distribution of polarisation ($   {P}$) {with respect to} position angle of {B-field}  ($\theta_{{B_{\rm pos}}}$) in the context of dust grain alignment with B-fields as shown in the left panel of Figure \ref{Fig:JHK}. The Gaussian fitted histogram of $\theta_{{B_{\rm pos}}}$ is shown in the right panel of the same figure for the corresponding bands and wavelength to illustrate the dispersion. Generally, a negative correlation trend is seen between the polarisation percentage, $   {P}$(\%), and the dispersion in $\theta_{{B_{\rm pos}}}$ because P(\%) is sensitive to the alignment efficiency of dust grains with respect to the B-field \citep{2021MNRAS.503.5006S}. {This is because polarisation tends to be highest where the dust grains are most aligned with the B-field  (near the mean ${\theta_{B_{\rm pos}}}$) and decreases symmetrically on either side as the dispersion increases, which corresponds to reduced dust alignment efficiency.} However, optical analysis shows no significant correlation between the two quantities, with r = -0.03, indicating that $   {P}$(\%) does not depend much on $\theta_{B_{\rm pos}}$ (see upper left panels of Figure \ref{Fig:JHK}). In contrast to the optical results, $   {P}$(\%) is observed to decrease with increasing dispersion in $\theta_{{B_{\rm pos}}}$ for the J, H, and Ks bands and at sub-mm wavelength (see the left panels of Figure \ref{Fig:JHK}, except for the upper panel),   {with correlation coefficients of r = -0.38, r = -0.44, r = -0.50 and r = -0.15, respectively.}

The optical data shows a maximum polarisation percentage of 6.2 $\pm$ 0.2\% and an average percentage of 1.8\%. However, the diffuse envelope of L328 core (NIR region) exhibits higher polarisation levels, with maximum values of 20.2~$\pm$~4.0\%, 9.2 $\pm$ 3.0\%, and 3.3 $\pm$ 0.9\%  and weighted average value of 10.75\%, 5.03\%, and 2.2\% in J, H and {Ks} bands, respectively. This result agrees with what we predicted by seeing vector plots in NIR region that the polarisation {is greater in the J-band than in the H-band, and greater in the H-band than in the Ks-band (see Figure \ref{fig:com_JHK})}. The dense core (at 850 $\mu$m) demonstrates a maximum polarisation of 19.84 $\pm$ 8.44\%.
 

\subsection{Magnetic field strength}{\label{mag_strength}}
We used the modified Davis-Chandrasekhar-Fermi (DCF; \citealt{1951PhRv...81..890D,1953ApJ...118..113C,2004ApJ...600..279C}) relation to estimate the {B-field} strength in the core of L328 using sub-mm polarisation measurements taken in this work and available molecular line observations from \cite{lee2013early}.

The DCF method determines the ﬁeld strength using the following equation
\begin{equation}
{B_{\rm pos}} = Q_{c} \sqrt{(4\pi\rho)}  \frac{\sigma_{\text{v}}}{\delta{\theta}} \,\, ,
\label{eq:OC}
\end{equation}
where $   {Q_{c}}$ is a correction factor, taken as 0.5, calculated from the simulations of turbulent clouds by \cite{2001ApJ...546..980}. $   {Q_{c}}$ accounts for variations of the B-field on scales smaller than the beam (\textit{i.e.}, more complex {B-field} and density structure that exist at smaller scales \citep{lai2001interferometric}), $\rho$ is the gas density defined as $\rho$ = $\mu_{g}{m_{{\rm H}}n({\rm H_{2}})}$ { with $\mu_{g}$ taken as 2.8,} and {${n(\rm H_{2})}$ is the number density of molecular hydrogen in cm$^{-3}$.} $\sigma_{\text{v}}$ is the average line-of-sight non-thermal velocity dispersion.
$\delta\theta$ is the dispersion in position angle and determines the local turbulence disrupting the B-field structure. The DCF method assumes that the geometry of the {B-field} is uniform, and so the dispersion of position angles is not greater than 25$^{\circ}$ {and the dispersion can be approximated as the standard deviation of the distribution}. We estimated the $\delta{\theta}$ from the observed errors and standard deviations of the measured polarisation position angles \citep{lai2001interferometric}. 

\begin{equation}
\delta{\theta} = \sqrt{\Delta\theta^{2}-\sigma_{\theta}^{2}} = \sqrt{22.6^{2}-9.7^{2}} = 20.4^{\circ}\,,
\end{equation}
where $\Delta\theta$ is the standard deviation in the distribution of the observed polarisation angles and $\sigma_{\theta}$ is the mean of measurement uncertainty of the polarisation angles. The $\delta{\theta}$ is calculated to be 20.4° $\pm$ 2.2° where the uncertainty in the dispersion angle is calculated by considering the standard deviation in the distribution of the uncertainties in the polarisation angles.
Now using ${n({\rm H_{2}})}$ and the non-thermal velocity ($\Delta$v$_{{{\rm NT}}}$, km s$^{-1}$), the DCF relation becomes
\begin{equation}
{B_{\rm pos}} \approx 9.3 \sqrt{{n({\rm H_{2}})}}  \frac{\Delta \text{v}_{{{\rm NT}}}}{\delta{\theta}} \mu {{\rm G}} \,\,,
\end{equation}
{and,}
\begin{equation}
\Delta \text{v} = \sqrt{\Delta \text{v}_{{{\rm T}}}^{2}+\Delta \text{v}_{{{\rm NT}}}^{2}}\,\,,
\end{equation}

where $\Delta$v is the total observed line-width, $\Delta$v$_{{{\rm T}}}$ is the thermal line-width, and $\Delta$v$_{{{\rm NT}}}$ is non-thermal line-width. The line-width $\Delta$v is related to velocity dispersion $\sigma_{\text{v}}$ as $\Delta$v = $\sigma_{\text{v}} \sqrt{ 8\, {\ln}\, 2}$.
\begin{equation}
{\sigma^2_{\text{v}_{\rm T}} = \text{v}^2_{{\rm sound}}} = \frac{k_{B}T_{{{\rm gas}}}} {\mu m_{{{\rm H}}}} \,\, , 
\end{equation}
where $\mu$ is the mean molecular weight of {gas}. The thermal velocity $\Delta$v$_{{{\rm T}}}$ is calculated as {0.14 km s$^{-1}$} by assuming the gas temperature is equal to dust temperature and using a molecular weight of 29 amu for N$_{2}$H$^{+}$ gas. The non-thermal component is used to approximate turbulent motion.

The total line-width ($\Delta$v) was calculated using the {N$_{2}$H$^{+}$} for sub-core S1, S2 and S3 as 0.5 $\pm$ 0.03 km s$^{-1}$, 0.61 $\pm$ 0.03 km s$^{-1}$ and 0.47 $\pm$ 0.02 km~s$^{-1}$, respectively, adopted from \cite{lee2013early}. The non-thermal component of velocity is calculated as 0.48, 0.59, and 0.45 km s$^{-1}$ for sub-core S1, S2, and S3, respectively, with an average value of 0.51 km s$^{-1}$.

The value of $n(\rm H_{2})$ was calculated as (4.7 $\pm$ 0.4) $\times$ 10$^{4}$ cm$^{-3}$, using the mass value of 0.69 $   {M_{\odot}}$, obtained from 850 $\mu$m dust continuum (see Sec. \ref{mass_est}).

The estimated $B_{\rm pos}$ for L328 core is found to be 50.5 $\mu$G. The {B-field} in its surrounding envelope was found to be $\sim$20 $\mu$G by \cite{soam2015first}. The uncertainty in the {B-field} strength was calculated using error propagation method as used by \cite{soam2019first} using the following relation

\begin{equation}\label{eq:dcferr}
 \frac{\delta {B_{\rm pos}}}{{B_{\rm pos}}} = \frac{1}{2}\frac{\delta n({\rm H}_{2})}{n({\rm H}_{2})} + \frac{\delta \Delta {\text{v}_{\rm NT}}}{\Delta {\text{v}_{\rm NT}}} + \frac{\delta(\delta_{\theta})}{\delta_{\theta}} \,\, ,
\end{equation}
where $\delta n({\rm H}_{2})$, $\delta \Delta {\text{v}_{\rm NT}}$, and $\delta(\delta_{\theta})$ are the uncertainties in $n({\rm H}_{2})$, {$\Delta \text{v}_{\rm NT}$}, and $\delta_{\theta}$, respectively.

The ${B_{\rm pos}}$ in the core is estimated as {50.5 $\pm$ 9.8} $\mu$G, which is approximately 2.5 times larger than the $B_{\rm pos}$ in the envelope {\citep[NIR region; ][]{soam2015first}},
indicating that the strength of {B-field} is higher in the core. This may be due to core collapse and enhanced magnetic flux in the core.

{Furthermore, \cite{2021A&A...647A.186S} proposed a modified method to estimate the B-field strength by assuming that all magnetohydrodynamics (MHD) modes are excited, including fast and slow modes. The B-field strength can be calculated using their equation}
\begin{equation}
{B_{\rm pos}} = \sqrt{{2\pi\rho}}  \frac{{\sigma_{\text{v}}}}{\sqrt{{\delta{\theta}}}} \,\,.
\label{eq:ST}
\end{equation}

This equation is the modified version of Equation \ref{eq:OC}, multiplied by ${\sqrt{2\times\delta\theta}}$, where ${\delta\theta}$ is in radians. Using this method, the calculated B-field strength is 42.58 ${\mu}$G. Overall, the Skalidis-Tassis method gives a smaller B-field strength compared to the  Ostriker-Crutcher DCF method.

\subsection{Mass-to-flux Ratio}

Now that we have {B-field} strength, we can test the relative importance of {B-fields} over gravity by {calculating the} mass-to-flux ratio, which is represented by a parameter $\lambda$ \citep{2004Ap&SS.292..225C}. The parameter $\lambda$ serves as an indicator of the balance between magnetic support and gravitational collapse in a given structure. When $\lambda$ < 1, the structure is considered ‘magnetically subcritical’, meaning it is supported against gravitational collapse by B-fields. Conversely, when $\lambda$ > 1, the B-field is insufficient to prevent gravitational collapse, and the structure is termed ‘magnetically supercritical’.

The observed B-field strength (50.5 $\pm$ 9.8 $\mu$G) and core column density {are used} in the estimation of $\lambda$.
 The $\rm H_{2}$ column density of L328 is found to be ${N({\rm H_{2}})}$ = {7.2 $\pm$ 0.6 $\times$ 10$^{21}$ cm$^{-2}$ using 850 $\mu$m continuum data by the given equation}

\begin{equation}
{N(\text{H}_2) = \frac{4}{3} n(\text{H}_2) \times r} \,\,,
\end{equation}
  {where $   {r}$ is the radius of core, with $   {r}$ = 36" (0.037 pc at a distance of 217 pc), consistent with 72" aperture used to estimate the L328 core, as described in the Sec. \ref{mass_est}. The $   {n}$(H$   {_{2}}$) is the volume density, with value adopted from previous section.}

The value of $\lambda$ is estimated using the the relation given by \citep{2004Ap&SS.292..225C}

\begin{equation}
\lambda = 7.6\times10^{-21} \frac{{N({\rm H_{2}})}/{{\rm cm}}^{-2}}{{B_{\rm pos}}/\mu {{\rm G}}}\,\,.
\end{equation}

The value of $\lambda$ comes out to be {1.1 $\pm$ 0.2}. This suggests that L328 core is magnetically transcritical. While calculating ${\lambda}$ from the mass estimation of 1.3 $   {M_{\odot}}$, the magnetic energy increases by a factor of ${\sqrt{1.8}}$, and ${N({\text{H}_2})}$ increases by a factor of ${1.8}$. Consequently, ${\lambda}$ will increase by a factor of ${\sqrt{1.8}}$, leading to a final ${\lambda}$ value of ${1.5}$, further showing it is magnetically supercritical.

In the envelope (NIR region) of L328, the value of $\lambda$ was found to be 1.3 $\pm$ 0.6 suggesting it to be marginally magnetic supercritical \citep{soam2015first}.

\subsection{Energy budget of the core}

To evaluate the energy budget of the core, we used sub-mm data, as it traces the B-field lines within the core.
The energy budget of L328 core is studied by comparing magnetic, kinetic (both thermal and non-thermal), and gravitational energies.
The magnetic energy can be calculated using the equation

\begin{equation}
{E_{\rm mag}}=\frac{{B^{2}_{\rm total}}V}{{8\pi}} \,\,,
\end{equation}
where ${E_{\rm mag}}$ is the total magnetic energy,  \textit{V} is the core volume   $(= 4/3\times \pi r^3)$ with   $r$ as the radius of core, and ${B_{\rm total}}$
 is total {B-field} strength. The total {B-field} is the sum of plane of sky {B-field} (${B_{\rm pos}}$) and the line of sight {B-field} that is observed by Zeeman effect. Since no Zeeman observation is done for L328 core, we considered two cases: 1) ${B_{\rm total}}$ = ${B_{\rm pos}}$, and 2) ${B_{\rm total}}$ $\approx$ 1.3 $\times$ ${B_{\rm pos}}$, as proposed by \cite{2004ApJ...600..279C} who suggested that the strength of 3-dimensional {B-field} (${B_{\rm total}}$) is related to plane of sky {B-field} (${B_{\rm pos}}$) as ${B_{\rm pos}}$ = $\frac{\pi}{4}$ $\times$ ${B_{\rm total}}$.
In the former case, the total magnetic energy is calculated as 5.7 $\times$ 10${^{41}}$ ergs, and in the latter case, it is 9.6 $\times$ 10${^{42}}$ ergs.

The non-thermal kinetic energy can be calculated as {}
\begin{equation}
{E_{\rm NT, \, kin}}=\frac{3M\sigma^{2}_{\text{v}, \,{{\rm NT}}}}{2}\,\,,
\end{equation}
where   {\textit{M}} represents mass of the core and $\sigma_{\text{v}, \, {\rm NT}}$ denotes non-thermal velocity. The calculated non-thermal kinetic energy is 9.8 $\times$ 10${^{41}}$ {ergs}.
{The ratio of total turbulent energy and magnetic energy is 2.5 and 1.5 in the former and latter cases, respectively.}

The thermal kinetic energy can be calculated as
\begin{equation}
{E_{\rm T, \, kin}}=\frac{3M\sigma^{2}_{\text{v}, \, {{\rm T}}}}{2}\,\,,
\end{equation}
where   {\textit{M}} represents mass of the core and $\sigma_{\text{v},\, {{\rm T}}}$ denotes thermal velocity. The calculated thermal kinetic energy is 7.4 $\times$ ${10^{40}}$ {ergs}.

The gravitational energy for L328 core by considering it as a sphere can be calculated as 
\begin{equation}
    E_{g}= -\frac{3GM^2}{5R}\,\,.
\end{equation}
The calculated gravitational energy is {6.9 $\times$ 10${^{41}}$} {ergs}.

\begin{table*}
\centering
\caption{{Energy budget of the L328 core, showing values for two mass estimates.}}
\label{tab:energy}
\begin{tabular}{lcccc} 
\hline
{Mass ($   {M}_{{\odot}}$)} & {{$E{_{\rm mag}}$ $(\times 10^{{41}}$ ergs)}} & {{$E{_{\rm NT, \, kin}}$ $(\times 10^{{41}}$ ergs)}} & {{$E{_{g}}$ $(\times 10^{{41}}$ ergs)}} & {$E{_{\rm T, \, kin}}$ $(\times 10^{{41}}$ ergs)}\\

\hline
{0.69} & {5.7} & {9.8} & {6.9} & {0.74} \\
{1.3}  & {5.7} & {18}  & {24}  & {1.4} \\
\hline 
\end{tabular}
\begin{flushleft}
{Note: The comparison of these energies highlights the roles of {B-fields}, turbulence, and gravity in the core’s dynamics and stability.}
\end{flushleft}
\end{table*}

Table \ref{tab:energy} summarizes the energy distribution in the L328 core for two different mass estimates. In both cases, the thermal kinetic energy is significantly lower than the  gravitational energy, suggesting a tendency for the core to collapse. However, the presence of significant magnetic and turbulent energies indicates that these forces may counteract the gravitational pull, potentially delaying or regulating the collapse process within the core. In the first case, the magnetic, non-thermal and gravitational energies are comparable to each other. In the other case, gravitational energy is approximately 4 times and turbulent energy is 3 times greater than magnetic energy, indicating a stronger influence of gravity and turbulence in this scenario.

\subsection{Polarisation hole}

\begin{figure*}
\resizebox{12.0cm}{8.0cm}{\includegraphics{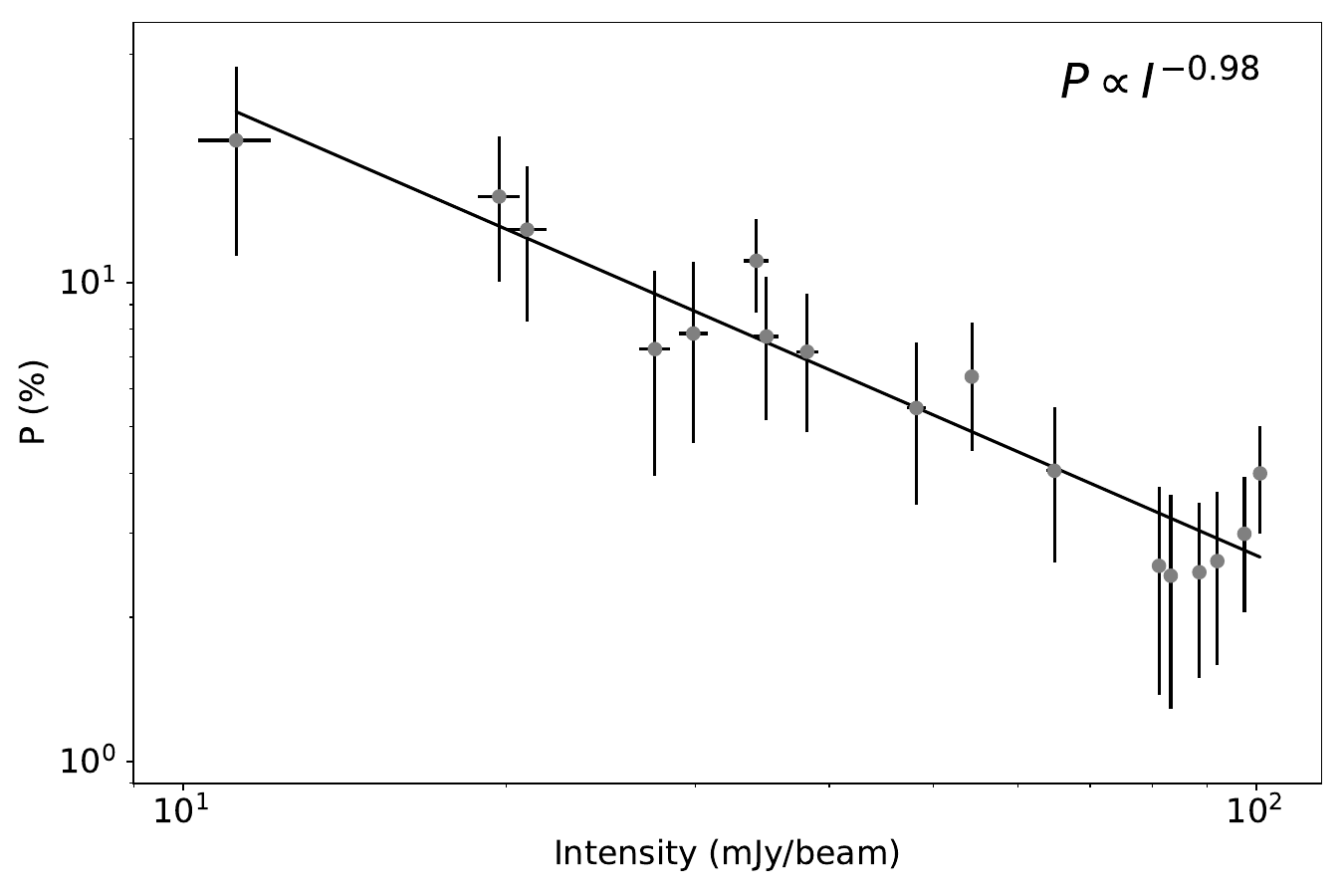}}
\caption{Polarisation fraction variation with intensity in the L328 core, based on values and uncertainties from POL-2 measurements.}\label{Fig:Power_law}
\end{figure*}

We investigated the variation in polarisation fraction from diffuse to high-density regimes (as we go radially inward) in the L328 core. These kind of investigations have already been done in several other cores {\citep{2000ApJ...531..868M,lai2001interferometric, 2018A&A...620A..26J, 2018ApJ...855...39K, soam2019first}}. The L328 core shown in the lower left panel of Figure \ref{Fig:manetic+pol} shows the polarisation vectors with their length depending on the degree of polarisation. It is clear that the length of vectors towards the higher-density parts is smaller than those of the vectors lying on the core boundary. This indicates a drop in polarisation fraction towards higher densities. This effect is known as 'depolarisation' and can be further analysed quantitatively by comparing the degree of polarisation with total intensity using the relation, $   {P}$ $\propto$ $   {I}$ $^{-\alpha}$.


We plotted the 850 $\mu$m polarisation versus the intensity in Figure \ref{Fig:Power_law}. The figure shows a negative correlation between $   {P}$ and $   {I}$ with a slope of $\alpha$ = 0.98 $\pm$ 0.08 that is consistent with the polarisation hole seen in L1521F, a core with a VeLLO in Taurus studied by \cite{soam2019first}.  This core is similar to L328 core and also has three sub-cores detected at 870 $\mu$m \citep{tokuda2016revealing}. There are accepted reasons for 'depolarisation' seen in starless and star-forming cores. One possible reason is changes in B-field orientation in the denser regions. The grain growth in the dense cold regions of the cores can also contribute to the drop in polarisation because the bigger grains becoming more spherical will no longer be efficiently aligned with the B-fields. As smaller grains coagulate and form larger aggregates, they tend to become more spherical due to surface energy minimization \citep{ALV2015}. Additionally, in dense regions, gas randomisation resulting from gas-grain collisions further
disrupts the alignment of dust grains, decreasing alignment efficiency and leading to lower polarisation \citep{2007JQSRT.106..225L, 2021ApJ...907...93S}. Magnetic reconnection is also explored as one of the possible reasons  \citep{1999ApJ...517..700L}. It occurs when B-field lines break and reconnect, disrupting the uniformity of the field. Since dust grains align with the local B-field, this disruption reduces their alignment efficiency. Additionally, insufficient radiation causing weak RAT \citep{2007MNRAS.378..910L} is another explanation for depolarisation.

\begin{table*}
	\centering
	\caption{The comparison of magnetic energy with other cores.}
	\label{tab:compare}
	\begin{tabular}{lcccr} 
		\hline
            Name & IRAS source/starless & Temperature &Mass-to-flux ratio & magnetically  \\ 
                 &                  &   T (K)    &      & subcritical/supercritical \\  
             \hline
            L1521F & IRAS & 10 & 3.1$\pm$0.2 &  supercritical\\
            L328   & IRAS &{11.5} & {1.1$\pm$0.2} & supercritical\\
            L183   & starless & 8.5 & 0.26$\pm$0.14 & subcritical\\
            L1512  & starless & 7.5&1.2$\pm$0.8 & slightly supercritical\\
            {L1544} & {starless} & {10} & {0.8} & {slightly subcritical}\\
            \hline 
          
	\end{tabular}
    \begin{tablenotes}
     \item  Note: The values for L1521F, L183, L1512 and L1544 are taken from \cite{fukaya2023twisted}, \cite{karoly2020revisiting}, \cite{2024ApJ...961..117L}, {and \citep{1999MNRAS.305..143W,2004ApJ...600..279C},} respectively.
    \end{tablenotes}
    
\end{table*}

\subsection{Comparison with other studies}
\cite{kim2016search} reported 95 VeLLOs in {the} Gould Belt but only few have been studied for the sub-mm polarisation. Table \ref{tab:compare} presents a comparative analysis of selected VeLLO cores with other starless, {chemically-evolved} cores to discern the factors contributing to their starless nature despite being chemically evolved. L1512 is a starless core but it is chemically evolved {\citep{lin2020physical}, and recently, in \cite{2024ApJ...961..117L},} they suggest that L1512 may have just recently reached supercriticality and will collapse at any time (highlighting the dynamic nature of such cores). Similarly, L1544, another starless yet chemically-evolved core, shows infall signatures and a marginally subcritical mass-to-flux ratio \citep{2004ApJ...600..279C}, suggesting that it is likely to collapse under its gravity, leading to star formation  \citep{2005ApJ...619..379C}. On the other hand, L183 is also chemically evolved \citep{tafalla2005observations} but remains starless. In contrast, L328 is not chemically-evolved but still has an IRS source (i.e., VeLLO),
{indicating that despite lacking the chemical evolution of cores like L1512 and L1544, it still harbors an embedded protostar.}

A common thread emerging from this comparison is the potential link between magnetic supercriticality and the presence of IRS sources or imminent collapse. Cores exhibiting magnetic subcriticality, on the other hand, tend to remain starless \citep{karoly2020revisiting}. However, the limited dataset underscores the necessity for an expanded sub-mm polarisation study on VeLLOs to draw definitive conclusions regarding the intricate interplay of {B-field}, chemical evolution, and the star formation process. Further investigations in this direction promise to unveil the underlying mechanisms governing the diverse outcomes observed in VeLLOs. The available data is insufficient for making definitive statements; additional sub-mm polarisation studies on VeLLOs and cores are imperative to draw meaningful conclusions.

\begin{table*}
	\centering
	\caption{Results of JCMT/POL-2 Observations of L328 core at 850 $\mu$m Wavelength}
	\label{tab:vector}
	\begin{tabular}{lccccr} 
		\hline
            ID & $\alpha$(J2000) & $\delta$(J2000) & $I \pm \sigma_{I}$ & P $\pm \sigma_{P}$ & $\theta\pm\sigma_{\theta}$ \\ 
               &                 &                 & (mJy/beam) &    ($\%$)       &      ($\deg$)   \\  
            
            \hline 
            1 & 18:16:59.220 & -18:02:38.5 & 48.22 $\pm$ 0.98 & 5.47 $\pm$ 2.03 & 26.3 $\pm$ 9.84\\ 
            2 &  18:17:01.744 & -18:02:26.5 & 19.71 $\pm$ 0.87 & 15.14 $\pm$ 5.1 & 38.59 $\pm$ 8.43\\ 
            3 &  18:17:00.903 & -18:02:26.5 & 54.3 $\pm$ 0.81 & 6.37 $\pm$ 1.91 & 34.18 $\pm$ 7.27\\ 
            4 &  18:17:00.062 & -18:02:26.5 & 83.21 $\pm$ 0.88 & 2.44 $\pm$ 1.16 & 178.79 $\pm$ 11.67\\ 
            5 &  18:16:59.220 & -18:02:26.5 & 88.48 $\pm$ 0.87 & 2.48 $\pm$ 0.99 & 10.53 $\pm$ 10.92\\ 
            6 &  18:16:58.379 & -18:02:26.5 & 34.22 $\pm$ 0.93 & 11.12 $\pm$ 2.47 & 39.81 $\pm$ 6.28\\ 
            7 &  18:16:59.220 & -18:02:14.5 & 97.45 $\pm$ 0.97 & 2.99 $\pm$ 0.94 & 21.15 $\pm$ 8.62\\ 
            8 &  18:16:58.379 & -18:02:14.5 & 38.15 $\pm$ 0.9 & 7.18 $\pm$ 2.31 & 54.46 $\pm$ 9.43\\ 
            9 &  18:17:03.427 & -18:02:02.5 & 11.21 $\pm$ 0.87 & 19.84 $\pm$ 8.44 & 13.24 $\pm$ 11.63\\ 
            10 &  18:17:02.586 & -18:02:02.5 & 29.88 $\pm$ 0.93 & 7.83 $\pm$ 3.21 & 6.99 $\pm$ 11.63\\ 
            11 &  18:17:01.744 & -18:02:02.5 & 81.14 $\pm$ 0.96 & 2.56 $\pm$ 1.19 & 30.86 $\pm$ 12.06\\ 
            12 &  18:16:59.220 & -18:02:02.5 & 100.79 $\pm$ 1.04 & 4.0 $\pm$ 1.01 & 43.21 $\pm$ 6.78\\ 
            13 &  18:17:00.062 & -18:01:50.5 & 91.95 $\pm$ 0.81 & 2.62 $\pm$ 1.03 & 94.79 $\pm$ 10.46\\ 
            14 &  18:16:59.220 & -18:01:50.5 & 64.83 $\pm$ 1.07 & 4.05 $\pm$ 1.45 & 59.7 $\pm$ 9.78\\ 
            15 &  18:17:00.903 & -18:01:38.5 & 20.92 $\pm$ 0.88 & 12.91 $\pm$ 4.62 & 114.44 $\pm$ 9.6\\ 
            16 &  18:17:00.062 & -18:01:38.5 & 34.97 $\pm$ 0.94 & 7.72 $\pm$ 2.57 & 79.29 $\pm$ 8.93\\ 
            17 &  18:16:59.220 & -18:01:38.5 & 27.53 $\pm$ 0.93 & 7.27 $\pm$ 3.32 & 50.66 $\pm$ 12.36\\
            
            \hline
	\end{tabular}
    \begin{tablenotes}
     \item   Note: The $\theta$ is rotated by 90° to trace {B-field}.
    \end{tablenotes}
    
\end{table*}

\section{Summary}\label{summary}

The paper presents an observational study on various scales by utilizing data from $Planck$ ({at 850 $\mu$m)}, optical, NIR, and sub-mm dust polarisation, to trace parsec to sub-parsec scale B-fields. The key findings are summarized as follows:

\begin{itemize}

    \item The mass of L328 core and its sub-cores (S1 and S2) is found to be {0.69 $   {M_{\odot}}$, 0.34 $   {M_{\odot}}$, and 0.08$   {M_{\odot}}$,} respectively.
    
    \item We found that the cloud{-}scale {B-field} is well{-}connected to core{-}scale {B-field} by showing {overall orientation in  {Northeast-Southwest} direction}. This also indicates that the core is embedded in the strong {B-field} region.

    \item The {B-field} strength within the L328 core is estimated to be $\approx${50.5 $\pm$ 9.8 $\mu$G}, significantly higher (greater than {2.5} times) than the estimated value in the envelope. The core and envelope are found to be {transcritical} and marginally supercritical with a $\lambda$ value of {1.1 $\pm$ 0.2} and 1.3 $\pm$ 0.6, respectively.

    \item  {The non-thermal kinetic energy, gravitational and magnetic energies, are comparable to each other, while the thermal kinetic energy within the core are significantly less than the other three energies.}
    
    \item The polarisation fraction as a function of total intensity is found to be decreasing in the high-density region, indicating depolarisation in the core with a power-law slope of $\alpha$ = -0.98.

\end{itemize}

\section*{Acknowledgements}
We thank the anonymous referee for a thorough reading of our manuscript and giving us very useful comments that have significantly improved the quality of our paper.

This research has made use of the SIMBAD database, operated at CDS, Strasbourg, France. We also acknowledge the use of NASA’s SkyView facility {(\href{http://skyview.gsfc.nasa.gov}{http://skyview.gsfc.nasa.gov})} located at NASA Goddard Space Flight Center. J.K. acknowledges the Moses Holden Fellowship in support of his PhD and is now supported by the Royal Society under grant number RF\textbackslash ERE\textbackslash231132, as part of project URF\textbackslash R1\textbackslash211322. C.W.L. was supported by Basic Science Research Program through the National Research Foundation of Korea (NRF) funded by the Ministry of Education, Science, and Technology.
The JCMT is operated by the East Asian Observatory on behalf of National Astronomical Observatory of Japan; Academia Sinica Institute of Astronomy and Astrophysics; the Korea Astronomy and Space Science Institute; the Operation, Maintenance and Upgrading Fund for Astronomical Telescopes and Facility Instruments, budgeted from the Ministry of Finance of China. SCUBA-2 and POL-2 were built through grants from the Canada Foundation for Innovation. This research used the facilities of the Canadian Astronomy Data Centre operated by the National Research Council of Canada with the support of the Canadian Space Agency. SPIRE has been developed by a consortium of insti-
tutes led by Cardiff Univ. (UK) and including: Univ. Lethbridge (Canada);
NAOC (China); CEA, LAM (France); IFSI, Univ. Padua (Italy); IAC (Spain);
Stockholm Observatory (Sweden); Imperial College London, RAL, UCL-
MSSL, UKATC, Univ. Sussex (UK); and Caltech, JPL, NHSC, Univ. Colorado
(USA). This development has been supported by national funding agencies:
CSA (Canada); NAOC (China); CEA, CNES, CNRS (France); ASI (Italy);
MCINN (Spain); SNSB (Sweden); STFC, UKSA (UK); and NASA (USA).
PACS has been developed by a consortium of institutes led by MPE (Germany)
and including UVIE (Austria); KUL, CSL, IMEC (Belgium); CEA, OAMP
(France); MPIA (Germany); IFSI, OAP/AOT, OAA/CAISMI, LENS, SISSA
(Italy); IAC (Spain). This development has been supported by the funding
agencies BMVIT (Austria), ESA-PRODEX (Belgium), CEA/CNES (France),
DLR (Germany), ASI (Italy), and CICT/MCT (Spain). 

\section*{Data Availability}
The data used in this article will be shared on reasonable request to the corresponding author.

$Software$: Starlink \citep{currie2014}, SMURF \citep{chapin2013}, APLpy  \citep{2012ascl.soft08017R,aplpy2019}, Astropy \citep{astropy:2013,astropy:2018,astropy:2022}, SciPy \citep{2020SciPy-NMeth}, PyAstronomy \citep{2019ascl.soft06010C}, Numpy \citep{ harris2020array}.






\bibliographystyle{mnras}
\bibliography{refL328} 



\appendix

\bsp	
\label{lastpage}
\end{document}